\documentclass[aps,prl,reprint,amsmath,amssymb,groupedaddress,showpacs
]{revtex4-1}
\usepackage{graphicx}
\usepackage{dcolumn}
\usepackage{bm}
\usepackage{hyperref}
\usepackage{soul} 
\usepackage{color}
\begin{document}


\title{The ultimate absorption at light scattering by {a single obstacle}}

\author{Andrey E. Miroshnichenko}
\affiliation{Nonlinear Physics Centre, Australian National University, Acton, ACT 2601, Australia}
\email[
Corresponding author:\\ ]{\mbox{andrey.miroshnichenko\_at\_anu.edu.au};\\ replace ``\_at\_" by ``@"}
\author{Michael I. Tribelsky}
\email[E-mail: ]{tribelsky\_at\_mirea.ru; replace ``\_at\_" by ``@"}
\affiliation{Lomonosov Moscow State University, Moscow, 119991, Russia}
\affiliation{Moscow Technological University MIREA, Moscow, 119454, Russia}

\date{\today}

\begin{abstract}
{Based on fundamental properties of light scattering by {a particle} we reveal the existence of  {the} ultimate upper limit for the light absorption by any partial mode.  {First, we} obtain this result for  {scattering of a} plane wave by a symmetric spherical or infinite cylindrical structure of an arbitrary radius.  {Then, we generalize it to an arbitrary finite obstacle}. Importantly, the result is true for any polarization, any angle of incidence of the plane wave and any type of the structure (homogeneous, stratified, or with smoothly variable refractive index). The corresponding maximal partial cross-section is a universal quantity, which does not depend on the optical constants of the scatterer its radius, and even its shape.}
\end{abstract}


\maketitle
%
%
The problem of laser heating of absorbing particles in transparent media is important for various applications. Initially, the interest in this problem was related to the optical damage to transparent media initiated by such particles \cite{Manenkov,Anisimov_Book}. Then, the area of applications of the problem has expanded dramatically. Nowadays, it includes (but is not limited to) broadband solar light absorption~\cite{Kravets:PRB:2010,Subramania:PRL:2011}, cancer therapy and diagnosis \cite{cancer,Diagnostics}, microsurgery \cite{Microsurgery}, drug and gene delivery and release \cite{Delivery,Containers}, and nanoscale control of temperature distribution \cite{Plasmonics}. Local, nanoscale laser heating of magnetic materials beyond the Curie temperature may be employed for superhigh-density data recording and storage \cite{Recording}, etc.

In most of these problems the optimization of the heating of particles is highly desirable~\cite{I_EPL,WePRX,WeNJP}. The first step towards the optimization is the selection of the optical parameters of the scattering particle so that they maximize the absorption cross-section. A number of publications have been devoted to this important subject, see, e.g.,~\cite{WePRX,I_EPL,Thongrattanasiri:PRL:2012,Fleury:PRB:2014,Tretyakov,Asadchy:PRX:2015, Kivshar_arXiv,Miller:OE:2016}. Meanwhile, the fundamental question, whether there is the ultimate upper limit for the absorption in light scattering by an obstacle with a given size and shape, and if so, what is the value of this limit, still remains open.

{It should be emphasized that, while in 1D cases the absorbed power density in principle cannot exceed the incident one, 2D and 3D cases are qualitatively different. Specifically, in the 2D and 3D cases at optical resonances a scatterer acts as a funnel, collecting the incident radiation from a wide ``inlet" in the ``upstream" area and delivering the electromagnetic field to a small ``outlet" --- the proximity of the scattering obstacle~\cite{WeNJP,Bohren_AJP}. It may result in a huge enhancement of the field, concentrating within the scatterer and the power dissipated there. Thus, the answer to the question about the ultimate absorption in 2D and 3D is far from obvious.}

We should note, that certain specific cases related to this question have been discussed already. For example, in previous publication of one of the authors (MIT)~\cite{I_EPL} it has been shown that such a limit indeed exists at the scattering of a plane, linearly polarized electromagnetic wave by a small (relative to the wavelength of the incident light), spatially uniform plasmonic sphere. Similar results have been reported in Ref.~\mbox{\cite{Grigoriev:AP:2015}}. The corresponding maximal partial scattering cross-section is given by the following simple expression:
\begin{equation}\label{sigma_l_small}
  \sigma^{(\ell)}_{{\rm abs\; max}} = \frac{(2\ell + 1)\pi}{2 k^2},
\end{equation}
where $\ell$ stands for the multipolarity of the plasmonic resonance ($\ell = 1$ --- dipolar, $\ell = 2$ --- quadrupolar, etc.) and $k$ designates the wavenumber of the incident light in a vacuum. The maximal absorption is achieved at \emph{small} value of the imaginary part of the particle permittivity, when the partial scattering cross-section {becomes} equal to the absorption. Note, that the above expression for $\sigma^{(\ell)}_{{\rm abs\; max}}$ does not depend on the particle radius and its optical constants. An analogous conclusion may be made from the results of publication~\cite{Fan}, where the light scattering by an infinite circular cylinder with the perpendicular incidence of the scattered wave with the so-called TE polarization has been discussed.

{Note also the resent attempts to achieve the total absorption by a plasmonic nanoparticle irradiated by the adequately shaped incident beam~\cite{Sentenac:OL:2013}. A natural extension of this attempt to the irradiation by a plane incident wave would be the vanishing of the scattering cross-section, so that the extinction cross becomes equal to the absorption one. However, such an equality cannot be the case, since it contradicts to \emph{the optical theorem}, stipulating that the extinction cross-section is proportional to a certain amplitude component of the forward scattered wave~\cite{Kerker,Bohren_book}. Thus, any absorption is inevitably accompanied by some scattering. It makes the question about the ultimate absorption closely related to the scattering problem, see below.}

{In the present paper we reveal  {existence of}  {\it the ultimate  {upper} limit for the absorption  {of any} partial wave} for  {an} arbitrary finite 3D  {or 2D single} scatter for any polarization of the incident wave, its angle of incidence and material parameters of the scatterer. In particular, our results remain valid for multi-layered structures and the ones with a smoothly variable profile of the permittivity.
}

{To begin with, let us discuss the well known exact solutions for the light scattering by a homogeneous sphere and cylinder~\cite{Kerker,Bohren_book}. In the case of a sphere} the scattering cross-sections may be presented as a series of the so-called partial cross-sections:
\begin{eqnarray}
& & \sigma_{{\rm sca}}^{{\rm sph}} = \sum_{\ell = 1}^{\infty}\left\{\sigma_{{\rm sca}}^{{\rm sph} \; (a,\ell)} + \sigma_{{\rm sca}}^{{\rm sph} \; (b,\ell)}\right\} ;\label{sigma_sca_sph}\\
    & & \sigma_{{\rm sca}}^{{\rm sph} \; (a,\ell)} =\frac{2(2\ell+1) \pi}{k^2}|a_\ell|^2.\label{sigma_sca_a_part_sph}
\end{eqnarray}
The expression for $\sigma_{{\rm sca}}^{{\rm sph} \; (b,\ell)}$ is obtained from Eq.~\eqref{sigma_sca_a_part_sph} by the replacement $a \rightarrow b$.
Here $a_\ell$ and $b_\ell$ are the electric and magnetic complex scattering coefficients.

The absorption cross-section also may be presented as an analogous series with the partial cross-section
\begin{equation}\label{sigma_abs_a_part_sph}
  \sigma_{{\rm abs}}^{{\rm sph} \; (a,\ell)} =\frac{2(2\ell+1) \pi}{k^2}{\rm Re}\left(a_\ell - |a_\ell|^2\right),
\end{equation}
and $\sigma_{{\rm abs}}^{{\rm sph} \; (b,\ell)}$ following from $\sigma_{{\rm abs}}^{{\rm sph} \; (a,\ell)}$ at $a \rightarrow b$.

The case with a cylinder is more cumbersome. The point is that to find the solution of the diffraction problem  at an arbitrary angle of incidence of a plane, linearly polarized electromagnetic wave the incident radiation should be presented as a sum of the TE and TM modes. The same is true for the scattered field. As a result, instead of a single scattering and single absorption cross-section there are four pairs of them (for the cylinder they are related to a unit length of the cylinder) corresponding to the following partitions: \mbox{TE $\rightarrow$ TE}, \mbox{TE $\rightarrow$ TM}, \mbox{TM $\rightarrow$ TE} and \mbox{TM $\rightarrow$ TM}. It is convenient to write them as matrix entries, where indices 1 and 2 correspond to the TM and TE modes, respectively. Regarding the scattered coefficients, usually, to distinguish the notations for TM and TE modes from the multipolarity index $n$ instead of Arabic 1 and 2 Roman numerals $I$ and $I\!I$ are employed~\cite{Kerker}. Then, the diagonal and off-diagonal elements of the partial cross-sections are as follows:
\begin{eqnarray}
    \sigma_{11\;{\rm sca}}^{{\rm cyl}}& = &\frac{4 \pi}{k}\left\{ |b_{0I}|^2 + 2\sum_{\ell = 1}^{\infty}|b_{\ell I}|^2\right\};\label{sigma_sca11_cyl} \\
    \sigma_{12\;{\rm sca}}^{{\rm cyl}}& = &\frac{8 \pi}{k}\sum_{\ell = 1}^{\infty}|a_{\ell I}|^2;\label{sigma_sca12_cyl} \\
    \sigma_{11\;{\rm abs}}^{{\rm cyl}} &=& \frac{4\pi}{k}\left\{\vphantom{\sum_{\ell}^{infty}}{\rm Re}\left(b_{0I}-|b_{0I}|^2\right)\right. \nonumber\\
    & & +\left.2\sum_{\ell = 1}^{\infty}{\rm Re}\left(b_{\ell I}- |b_{\ell I}|^2\right)\right\}.\label{sigma_abs11_cyl}
 \end{eqnarray}
The other entries of matrices $ \sigma_{ij\;{\rm sca}}^{{\rm cyl}}$ and $ \sigma_{ij\;{\rm abs}}^{{\rm cyl}}$ look accordingly.

The scattering coefficients for the sphere are expressed in terms of the Riccati-Bessel functions. For the cylinder the Bessel functions are employed. {Even in the simplest case of a spatially homogeneous scatterer the expressions are cumbersome and will not be presented here.} For more details, see, e.g., books~\cite{Kerker,Bohren_book}.

A remarkable thing, however, is that \mbox{Eqs.~\eqref{sigma_sca_a_part_sph}, \eqref{sigma_sca11_cyl}--\eqref{sigma_sca12_cyl}}  have a similar universal structure. The same is true for Eqs.~\eqref{sigma_abs_a_part_sph}, \eqref{sigma_abs11_cyl}. {It should be stressed that Eqs.~\eqref{sigma_sca_sph}--\eqref{sigma_abs11_cyl} and others analogous for components of $ \sigma_{ij\;{\rm sca}}^{{\rm cyl}}$ and $ \sigma_{ij\;{\rm abs}}^{{\rm cyl}}$  are nothing but a direct consequence of the {optical theorem}~\cite{Kerker,Bohren_book}, presentation of the solution of the Maxwell equations in the series of spherical (cylindrical) harmonics and orthogonality of the harmonics with different values of $\ell$. The structure of these equations has {nothing} to do with the value and/or coordinate dependence of the complex permittivity of the scatterer $\epsilon = \epsilon' + i\epsilon''$~\cite{n1}, provided this dependence does not violate the spherical (cylindrical) symmetry of the one. Such a dependence, if any, affects just the values of the scattering coefficients. Therefore, Eqs.~\eqref{sigma_sca_a_part_sph}--\eqref{sigma_abs11_cyl} are valid for \emph{any} spherically (cylindrically) symmetric distribution of $\epsilon$, including its constant value (a spatially homogeneous scatterer); any smoothly variable $\epsilon(r)$; and a stratified scatterer, consisting of a number of layers with a stepwise variation of $\epsilon$ at transition from one layer to another.}


Thus, to maximize any partial absorption cross-section in a given scattering partition one has to find a maximum of the expression
\begin{equation}\label{z}
  {\rm Re}\left(z - |z|^2\right),
\end{equation}
see Eqs.~\eqref{sigma_abs_a_part_sph}, a single summand with a given value of $\ell$ in Eq.~\eqref{sigma_abs11_cyl} and the analogous expressions for $\sigma_{{\rm abs}}^{{\rm sph} \; (b,\ell)}$, $\sigma_{21\;{\rm abs}}^{{\rm cyl}},\; \sigma_{22\;{\rm abs}}^{{\rm cyl}}$, which are not presented here.

{As it has been mentioned above, the specific values of the scattering coefficients depend on the size parameter $q = Rk$, where R is the radius of the sphere (cylinder), and the profile $\epsilon(r)$. However, it is important that all these coefficients also have a universal structure and may be presented in the form:
%
%
\begin{equation}\label{FG}
  z=\frac{F}{F+iG} \equiv \frac{1}{1+i\zeta}.
\end{equation}
\mbox{}\\
\noindent
Here $\zeta \equiv G/F$. $F$ and $G$ are rather cumbersome functions~\cite{Kerker,Bohren_book}, whose explicit form is not required here.

{To prove the possibility of this presentation of $z$ in all the cases mentioned above we will follow the general arguments of Ref.~\cite{I_Coefficients}. Note that in the non-dissipative limit with $\epsilon'' = 0$ any partial absorption cross-section vanishes identically, i.e., $|z|^2 = {\rm Re}\,z$. Let $z=\mu+i\chi$. Then, the condition \mbox{$|z|^2 = {\rm Re}\,z$} may be rewritten as \mbox{$0 \leq \mu^2=\mu-\chi^2\leq \mu$.} It results in the constraint: $0 \leq \mu \leq 1$.}

{On the other hand, any real, nonnegative quantity, does not exceeding unity may be presented as the fraction
\begin{equation}\label{mu}
  \mu \equiv \frac{F^2}{F^2+G^2},
\end{equation}
where $F$ and $G$ are real. Eq.~\eqref{mu}, together with the equality $\chi^2 = \mu - \mu^2$, {which implies
\begin{equation}\label{chi}
  \chi \equiv \pm\frac{F G}{F^2+G^2},
\end{equation}
} bring about Eq.~\eqref{FG}.}
%

{However, the above proof in based on the vanishing of the partial absorption cross-section. For this reason it is valid in the non-dissipative case only. It may seem that at a finite $\epsilon''$ the numerator and/or denominator of Eq.~\eqref{FG} might have additional terms, violating this presentation.}

{If so, these terms should be proportional to a certain positive power of $\epsilon''$, since they must vanish at $\epsilon'' \rightarrow 0$. Now note that in the initial boundary value problem, describing the scattering, $\epsilon''$ enters only through $\epsilon = \epsilon'+i\epsilon''$, and any constructive solution of the problem determining the scattering coefficients does not separate $i\epsilon''$ from $\epsilon'$. Thus, at a transition from the non-dissipative limit to the case of finite dissipation the only way for $\epsilon''$ to enter Eq.~\eqref{FG} is to do that through the analytical continuation of real $\epsilon$ to the complex plane, i.e., through the formal replacement $\epsilon' \rightarrow \epsilon'+i\epsilon''$. It makes the functions $F$ and $G$ complex, but, naturally, does not violate the structure of Eq.~\eqref{FG}}.

{Of course, in any specific case one can arrive at the same conclusion by means of direct calculations, though they may be rather cumbersome. {An example of these calculations for the normal incidence of a plane wave on a stratified sphere can be found in Ref.~\cite{StratifiedSphere}.}}

Bearing all this in mind, Eq. \eqref{z} may be reduced to the following form:
\begin{equation}\label{xy}
   {\rm Re}\;z - |z|^2 =-\frac{y}{(1-y)^2+x^2},
\end{equation}
where $x ={\rm Re}\,\zeta,\; y = {\rm Im}\,\zeta$.

Since up to a certain positive prefactor Eq. \eqref{xy} equal to a partial absorption cross-section, which, by the definition, is a non-negative quantity, the first conclusion we can make is that \mbox{$y =$ Im$\;G/F \leq 0$.} The generic form of this restriction may be verified by direct calculations based upon the explicit expressions for $F$ and $G$ \cite{Kerker, Bohren_book}.

It is seen straightforwardly that Eq.~\eqref{xy} is maximized at $x=0$ and $y = -1$. The maximal value is 1/4. Note also, that at these values of $x$ and $y$ the following equality holds: $|z|^2 =  {\rm Re}\,z - |z|^2 = 1/4$, i.e., the partial absorption and scattering cross-sections equal each other.

It is easy to see that, actually, the partial absorption and scattering cross-sections equal each other at $y=-1$ and \emph{any} value of $x$. However, only at $x=0$ this equality holds at the point of the local maximum of the absorption cross-section.

{Another important point is that at $x=0$ and $y = -1$ $i\zeta = 1$, i.e., $z = 1/2$ being a purely real quantity. It provides the opportunity to detect the occurrence of the ultimate maximal absorption {just measuring the value} of the corresponding partial extinction or scattering cross-section together with the phase shift between the incident field and the one scattered by the obstacle to the corresponding multipolar partition.}

Collecting all together we may conclude that
\begin{itemize}
  \item For the problems in {question} the ultimate upper limit for a partial absorption cross-sections exists indeed.
  \item The corresponding values for a sphere equal the one, obtained earlier in Ref.~\cite{I_Coefficients,Grigoriev:AP:2015} for the case of a small plasmonic particle, see Eq.~\eqref{sigma_l_small}.
  \item For a cylinder the maximal partial absorption cross-section equals $\pi/k$ at $\ell = 0$ and double of that at any other values of $\ell$.
  \item To maximize the partial cross-section the two conditions: Re$(G/F) = 0$ and Im$(G/F) = -1$ must be satisfied simultaneously.
  \item If the conditions Re$(G/F) = 0$ and \mbox{Im$(G/F) = -1$} hold, the corresponding partial absorption and scattering cross-sections equal each other, while the complex scattering coefficient equals 1/2, becoming a purely real quantity. {In this case the partial \emph{scattering} cross-section for a sphere is also given by Eq.~\eqref{sigma_l_small}. For a cylinder it equals $\pi/k\;(\ell=0)$, or $2\pi/k\;(\ell \neq 0)$. The corresponding \emph{extinction} cross-section is double of these values.}
\end{itemize}

There are several important issues to be be clarified in view of these results. \emph{First}, the aforementioned analysis indicates that any partial absorption cross-section for a single scattering partition cannot exceed the obtained ultimate maximal values. However, it does not mean that these ultimate maxima may be always achieved. Just opposite --- as a rule for actual optical materials the two conditions Re$(G/F) = 0$ and Im$(G/F) = -1$ are inconsistent and cannot hold together. In these cases the actual maximal values of the partial absorption cross-section become smaller (sometimes much smaller) than the obtained ultimate limit.

\emph{Second}, intuitively it seems that the maximal absorption should happen close to the resonant frequencies, when the electromagnetic field, corresponding to the resonant mode, is maximal. In some cases this is true, indeed. For example, for a small ($q \ll 1$), weakly dissipating ($\epsilon'' \ll 1$) particle functions $F$ and $G$ are almost purely real~\cite{I_EPL}. Then, Re$(G/F) \approx {\rm Re}\,G/{\rm Re}\,F$ and the condition Re$(G/F) = 0$ is satisfied in the vicinity of the roots of equation ${\rm Re}\,G=0$ (function $F$ always remains finite). On the other hand, in this case the characteristic value of $F$ is small (e.g., for a sphere $F^{(a,\ell)} \sim q^{2\ell + 1},\;F^{(b,\ell)} \sim q^{2\ell + 3}$), while the characteristic value of $G$ is of the order of unity. Thus, the condition ${\rm Re}\,G=0$ approximately defines the poles of the scattering coefficients, i.e., the points of resonances, see Ref.~\cite{I_EPL} for more details.

However, such cases are rather exceptional. In a general, case functions $F$ and $G$ are of the same order of magnitude, and the roots of the equations Re$(G/F) = 0$ and Im$(G/F) = -1$ (if any) may have nothing to do with the resonant frequencies. It should also be stressed that when $F$ and $G$ are of the same order and the dissipation is not small, the definition of the resonances itself is not trivial and requires clarification.

To illustrate these general arguments the normalized partial cross-sections (efficiencies) for the electric dipole mode $Q_{{\rm sca}}^{{\rm sph} \; (a,1)} \equiv \sigma_{{\rm sca}}^{{\rm sph} \; (a,1)}/(\pi R^2)$ and $Q_{{\rm sca}}^{{\rm sph} \; (a,1)} \equiv \sigma_{{\rm abs}}^{{\rm sph} \; (a,1)}/(\pi R^2)$, calculated according to the exact Mie solution at $q=2$, are presented in Fig.~\ref{Fig:Q_a12_3D} as functions of $\epsilon'$ and $\epsilon''$. Mismatches between the positions of the crests in the two reliefs are seen clearly. Within the range of variation of ${\epsilon}$ presented in Fig.~\ref{Fig:Q_a12_3D} there are just two points of the maxima of the partial absorption cross-section, namely ${\epsilon} \approx 4.968 + i 1.362$ and ${\epsilon} \approx 14.780 + i 1.278$ (the latter approximately corresponds to permittivity of common semiconductor GaAs at the wavelength about 650~nm~\cite{RefractiveIndex}). The calculations show, that at both {these} points equality $\sigma_{{\rm abs}}^{{\rm sph} \; (a,1)}/(\pi R^2)=\sigma_{{\rm sca}}^{{\rm sph} \; (a,1)}/(\pi R^2) = 3/(2q^2)$, see Eq.~\eqref{sigma_l_small} at $\ell = 1$, holds with high accuracy, see Fig.~\ref{Fig:Q_a12_plane}.
\begin{figure}
  \centering
  \includegraphics[width=0.5\textwidth]{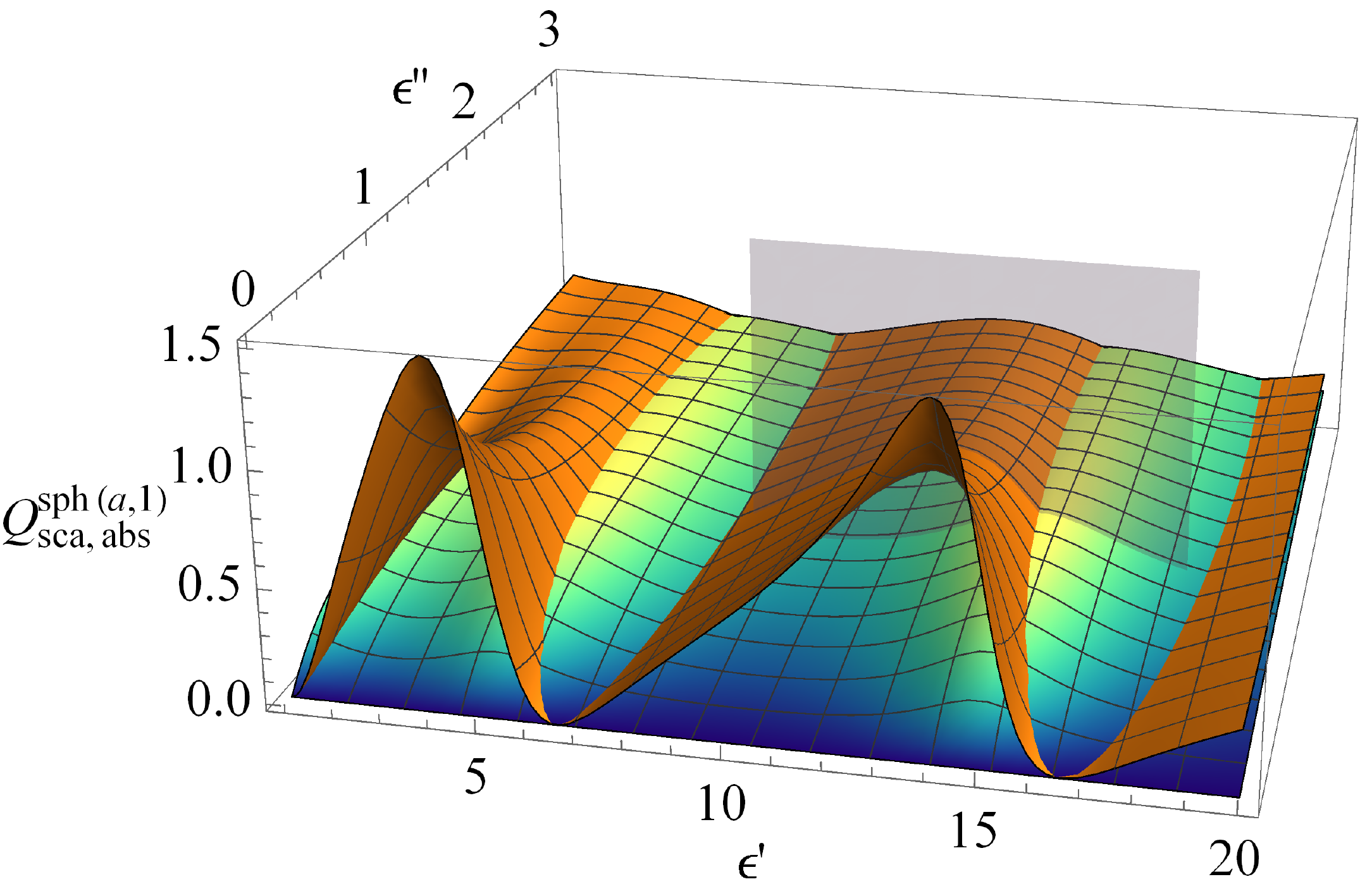}
  \caption{(color online) The exact Mie solution. Normalized partial scattering (brown) and absorption (blue-green-yellow) electric dipole cross-sections at light scattering by a sphere with $q=2$ as functions of its complex permittivity. Grey plane corresponds to $\epsilon'' = 1.278$. See the text for more details.}\label{Fig:Q_a12_3D}
\end{figure}
\begin{figure}
  \centering
  \includegraphics[width=0.45\textwidth]{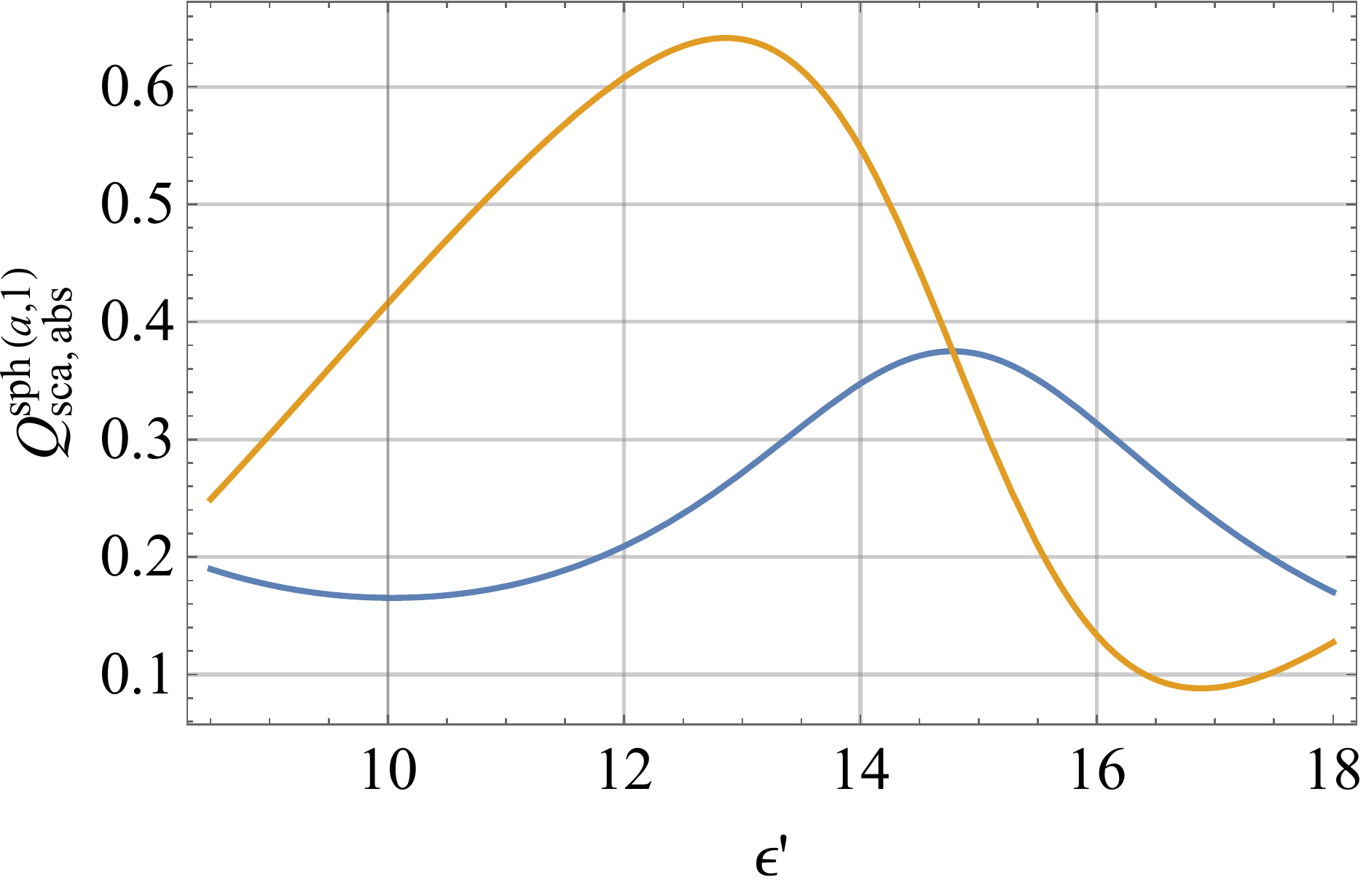}
  \caption{(color online) Cross-section of the plots shown in Fig.~\ref{Fig:Q_a12_3D} by the plane $\epsilon'' = 1.278$ (the grey plane in Fig.~\ref{Fig:Q_a12_3D}). Brown and blue lines correspond to $Q_{{\rm sca}}^{{\rm sph} \; (a,1)}$ and $Q_{{\rm abs}}^{{\rm sph} \; (a,1)}$, respectively. Note, the intersection of the two lines occurs exactly at the point of the local maximum of $Q_{{\rm abs}}^{{\rm sph} \; (a,1)}$. }\label{Fig:Q_a12_plane}
\end{figure}

It may be expected that the positions of the maxima of a partial cross-section are associated with those for the modulus of the amplitude of the corresponding partial mode excited \emph{within} the scatterer. However, the detailed discussion of this point lies beyond the scope of the present Letter and will be reported elsewhere, see also our recent paper~\cite{WearXiv}.

\emph{Third}, the obtained ultimate limits are the upper bounds for a partial cross-section for \emph{a single scattering partition}, while the net cross-section is a sum of an infinite number of the partial cross-sections, corresponding to each partition separately. If the maximization conditions  Re$(G/F) = 0$ and Im$(G/F) = -1$ may be satisfied for a number of partitions \emph{simultaneously}, it may result in a considerable enhancement of the net cross-section.

\emph{Fourth,} if a scatterer has additional variable parameters, it provides more freedom to tune the scatterer to the enhanced absorption. For example, 
a stratified scatterer, consisting of several layers of different materials with different thickness, obeys essentially the same rules. In this case tuning of the thickness of the layers and their permittivities may allow to achieve overlap of the maximal absorption, produced {by different modes, i.g., dipolar and quadrupolar,} and to build up a ``superabsorber"~\cite{Miroshn_particle,Miroshn_wire} in the same manner as it was suggested in Ref.~\cite{Fan} for a ``superscatterer".

{A few words about the partial \emph{scattering} and \emph{extinction} cross-sections should be added. Up to a ceratin positive prefactor the former and the latter equal Re$\,z$ and $|z|^2$, respectively~\cite{Kerker,Bohren_AJP,I_Coefficients}, where $z$ is defined according to Eq.~\eqref{z}. Bearing in mind that Im$\,\zeta \leq 0$, it is easy to see that both the quantities are maximized at $\zeta =0$, i.e., at $G=0$, If this condition holds, $z=1$ and the scattering cross-section equals the extinction one. Obviously, it can happen in the non-dissipative limit only (i.e., only in this limit the equation $G=0$ has physically meaningful roots). In other words, the ultimate maximal values for the partial extinction and scattering cross-sections are realized at $\epsilon'' = 0$ (see Fig.~\ref{Fig:Q_a12_3D}) at the resonances corresponding to the anomalous scattering~\cite{I_JTEP,I-Luk_PRL}}.

{What happens in the case of an arbitrary scatterer, which does not possess the spherical, or cylindrical symmetry? The multipolar expansion for this case is discussed in Ref.~\cite{Grahn}. The only difference between this general case and symmetrical, discussed above is that violation of the symmetry removes the degeneracy of the expansion in spherical (cylindrical) harmonics, so that it occurs of the general type, i.e., the sums in Eqs.~\eqref{sigma_sca_sph}, \eqref{sigma_sca11_cyl} are transformed as follows:
\begin{equation*}
  \sum_{\ell = 1}^{\infty} \rightarrow \sum_{\ell = 1}^{\infty}\sum_{m = -\ell}^{\ell}.
\end{equation*}
All the rest remains the same. Obviously this transformation does not affect the general reasoning presented above.}

{Thus, we have revealed a new fundamental feature of the light scattering problem. Based on the optical theorem, which relates the extinction, scattering and absorption cross-sections, we have proven \emph{rigorously}  the existence of the ultimate upper limit for the partial absorption cross-section for every individual scattering mode {in the multipolar expansion} for {\it an arbitrary scattering object}. In particular, we demonstrate our results for light scattering by a specially uniform sphere or cylinder of arbitrary sizes and internal structure. We have obtained the conditions for the limit to be achieved and the simple explicit expressions for the values of the partial absorption cross-sections at this limit. It occurs that this value is a fundamental quantity, which does not depend on the optical properties of the scatterer and/or its size. We believe the results shed a new light on this important problem and may stimulate further study of the phenomenon.}

The authors are grateful to Boris S. Luk'yanchuk for the fruitful discussions and valuable comments.
The work of AEM was supported by the Australian Research Council via the Future Fellowship program (FT110100037).


\begin{thebibliography}{99}
\bibitem{n1}If a scatterer is embedded into a transparent medium with refractive index $n^2$, $\epsilon$ stands for the normalized (over $n^2$) permittivity.
\end{thebibliography}
\end{document}